\documentclass{sig-alternate}
\makeatletter
\let\@copyrightspace\relax
\makeatother

\usepackage{color}
\usepackage{ntheorem}
\usepackage[caption=false]{subfig}

\newtheorem{definition}{Definition}[section]

\newtheorem{example}{Example}[section]
\newcommand{\eat}[1]{}

\begin{document}
\title{Bayesian Data Cleaning for Web Data}

\numberofauthors{1}
\author{
\alignauthor
 Yuheng Hu \quad
 Sushovan De \quad
 Yi Chen\quad
 Subbarao Kambhampati\thanks{This research is supported by ONR grant N000140910032
and two Google research awards} \quad\\
       \affaddr{Dept. of Computer Science and Engineering}\\
       \affaddr{Arizona State University}\\
       \email{\{yuhenghu, sushovan, yi, rao\}@asu.edu}\\
}

\maketitle

\begin{abstract}
Data Cleaning is a long standing problem, which is growing in
  importance with the mass of uncurated web data. State of the
  art approaches for  handling inconsistent data are systems that learn and use
  conditional functional dependencies (CFDs) to rectify data. These
  methods learn data patterns--CFDs--from a clean sample of the data
  and use them to rectify the dirty/inconsistent data. While getting a
  clean training sample is feasible in enterprise data scenarios, it
  is infeasible in web databases where there is no separate curated
  data. CFD based methods are unfortunately particularly sensitive to
  noise; we will empirically demonstrate that the number of CFDs
  learned falls quite drastically with even a small amount of noise.
  In order to overcome this limitation, we propose a fully
  probabilistic framework for cleaning data. Our approach involves
  learning both the generative and error (corruption) models of the
  data and using them to clean the data. For generative models, we
  learn Bayes networks from the data. For error models, we consider a
  maximum entropy framework for combing multiple error processes.
 The generative and error
  models are learned directly from the noisy data. We present the
  details of the framework and demonstrate its effectiveness in
  rectifying web data.

\end{abstract}

\section{Introduction}
\label{sec:introduction}
Real-world data is noisy and often suffers from corruptions that may
impact data understanding, data modeling and decision-making. This
situation is ubiquitous and even more severe when we deal with the web
data generated by users or automated programs. For example, humans can
introduce errors like \emph{typos} and \emph{omitted} data entries,
and automated approaches can introduce algorithmic errors such as
\emph{inaccurate} information extraction. Alleviating this problem
needs data cleaning, i.e., catching and fixing corruptions in the
data. In this paper, we focus on unsupervised cleaning for the
uncurated structured data on the web rife with incompleteness and
inconsistency.  By identifying and curing noisy values, it is possible
to gain deeper understanding of the data, improve models, or make
better decisions.

A variety of approaches have been proposed for data cleaning, from
traditional methods (e.g., outlier detection \cite{knorr2000distance},
noise removal \cite{xiong2006enhancing}, and imputation
\cite{fellegi1976systematic}) to recent effort on examining integrity
constraints, e.g., functional/inclusion dependencies (FD/IND)
\cite{bohannon2005cost} and their extensions
(CFD/CIND) \cite{fan2009discovering,leo2011cleaning}. Although these
methods are efficient in their own scenarios, they have severe
drawbacks when cleaning the noisy web data because: (1) State of the art approaches (e.g.,
\cite{bohannon2005cost,fan2009discovering,leo2011cleaning}) depend on
the availability of a clean data corpus or external reference table to
learn data quality rules/patterns before fixing the errors. Such clean
corpora may be easy to establish in a tightly controlled enterprise
environment but infeasible when on the web. One may attempt to learn
data quality rules directly from the noisy web data. Unfortunately, as
we will demonstrate in Section \ref{sec:cfd}, this attempt fails to
obtain any rules even with very small percentage of corruptions in the
data; (2) Many other approaches (e.g., \cite{knorr2000distance,xiong2006enhancing}) are
only concerned about identifying or removing noise (corruptions)
rather than fixing them; (3) Some of the prior work (e.g.,
\cite{kubica2002probabilistic,fellegi1976systematic}) only focuses on
fixing a single type of error. This is inadequate on the web where
multiple different kinds of corruptions could happen.

We answer the web data cleaning problem by devising an end-to-end
probabilistic framework on the available web data which involves
learning a model of the clean data generation process as well as an
error model of the corrupting process that introduces the noise. Then,
by treating the clean value as a latent random variable, our framework
leverages these two learned models and automatically infers its value
through a Bayesian estimation. There are several advantages to this
framework. First, modeling data probabilistically allows our framework
to tolerate possible noise in the training data\footnote{We assume
  only a small portion of the data is corrupted while the majority is
  clean.}. In other words, it relaxes the limiting requirement in the
existing approaches (i.e., building clean corpus in advance for
learning deterministic data quality rules). Second, explicitly and
naturally modeling the noise and the data corruption process through
an error model improves the accuracy and robustness of the noise
identification and fixing. For example, our error model can consider a
wide spectrum of errors that occur commonly on the web (e.g.,
misspelling, replacement and deletion errors). On the contrary, most
state-of-the-arts have to characterize each type of errors and develop
cleaning strategy for them separately. This is especially inconvenient
when a new type of error is found or the noisy data contains multiple
types of errors.

We evaluate the proposed framework rigorously on a real-world dataset
(used auto sales data). The results demonstrate the effectiveness and
efficiency of our method with respect to different sizes of the data
and various levels of noise in the data.

To summarize, our main contributions are as follows:
\begin{enumerate}
\item We find that although CFD-based approaches are designed to capture and fix dirty data, ironically, learning CFDs however depends on the availability of a perfectly or largely clean data corpus. Through the empirical experiment, we show that CFD learner fails to discover any CFDs from a dataset which contains a very small percentage of noise (0.1\%). Such discovery, to the best of our knowledge, has not been explored before, hence it is new.
\item We propose an end-to-end probabilistic framework for cleaning the dirty web data. Our approach involves
  learning both the data generative model and error (corruption) model from the input dirty dataset. For a possible corrupted tuple, our framework leverages these two learned models and then automatically infers its value.
\item We conduct extensive experiments to evaluate the effectiveness and efficiency for our framework. The experiments are performed on a real web dataset with different types and levels of corruptions introduced.
\end{enumerate}

The rest of the paper is organized as follows. In Section \ref{sec:related} we discuss related work. Section \ref{sec:cfd} presents the performance of CFD-based approaches on real noisy web data. In Section \ref{sec:approach} we present our approach. Quantitative evaluations are described in Section \ref{sec:experiments}. We conclude the paper in Section \ref{sec:conclusion}.

\section{Related Work}
\label{sec:related}
Recent years have witnessed a significant research interest in data cleaning and enhancing data quality. A variety of approaches have been proposed with focus on noise elimination, missing value prediction, and noisy value correction. Some of them work directly on detecting and removing data corruptions but without fixing them, such as outlier detection \cite{knorr2000distance} and noise removal \cite{xiong2006enhancing}. On the other hand, some focus on fixing those corruptions alone, such as value imputation \cite{fellegi1976systematic}. More recently, integrity constraints-based approaches have been proposed to capture and fix data corruptions such that the resulting database $D'$ is either consistent and minimally differs from original database $D$ or certain errors in $D$ get fixed. These methods heavily use the editing rules which are generated from the (conditional) functional dependencies ({CFD}s or {FD}s), (conditional) inclusion dependencies ({IND}s or {CIND}s) or matching dependencies ({MD}s) found from the data \cite{bohannon2005cost,fan2009discovering,leo2011cleaning, wu2004using}.

The focus of most of the above works is to improve the quality of the
data from a closed domain (e.g., census data or enterprise data)
with a single type of error (either incompleteness or
inconsistency). Therefore, it is not clear whether applying them to
the noisy data on the web will work. This is because of the openness
of the web where many kinds of errors may co-exist. Furthermore, most
integrity constraints-based approaches require rules which are
deterministic and carefully tuned. However, given the uncertainties
on the web, the performance of these approaches cannot be
guaranteed. More importantly, learning editing rules requires a clean
training corpus of high quality (an implicit assumption made in most
of the work in this line, see discussions in Section
\ref{sec:cfd}). However, such corpora are infeasible to acquire on the
web. To tackle these limitations, in this paper, we propose an
end-to-end probabilistic framework which is designed to handle the
data cleaning problem for the web data.
%
\section{Limitations of CFD-based Methods for Cleaning Web Data}
\label{sec:cfd}
\begin{figure}[!h]
\centering\vspace{-2mm}
\includegraphics[trim = 48mm 45mm 48mm 40mm, clip, width=80mm, keepaspectratio]{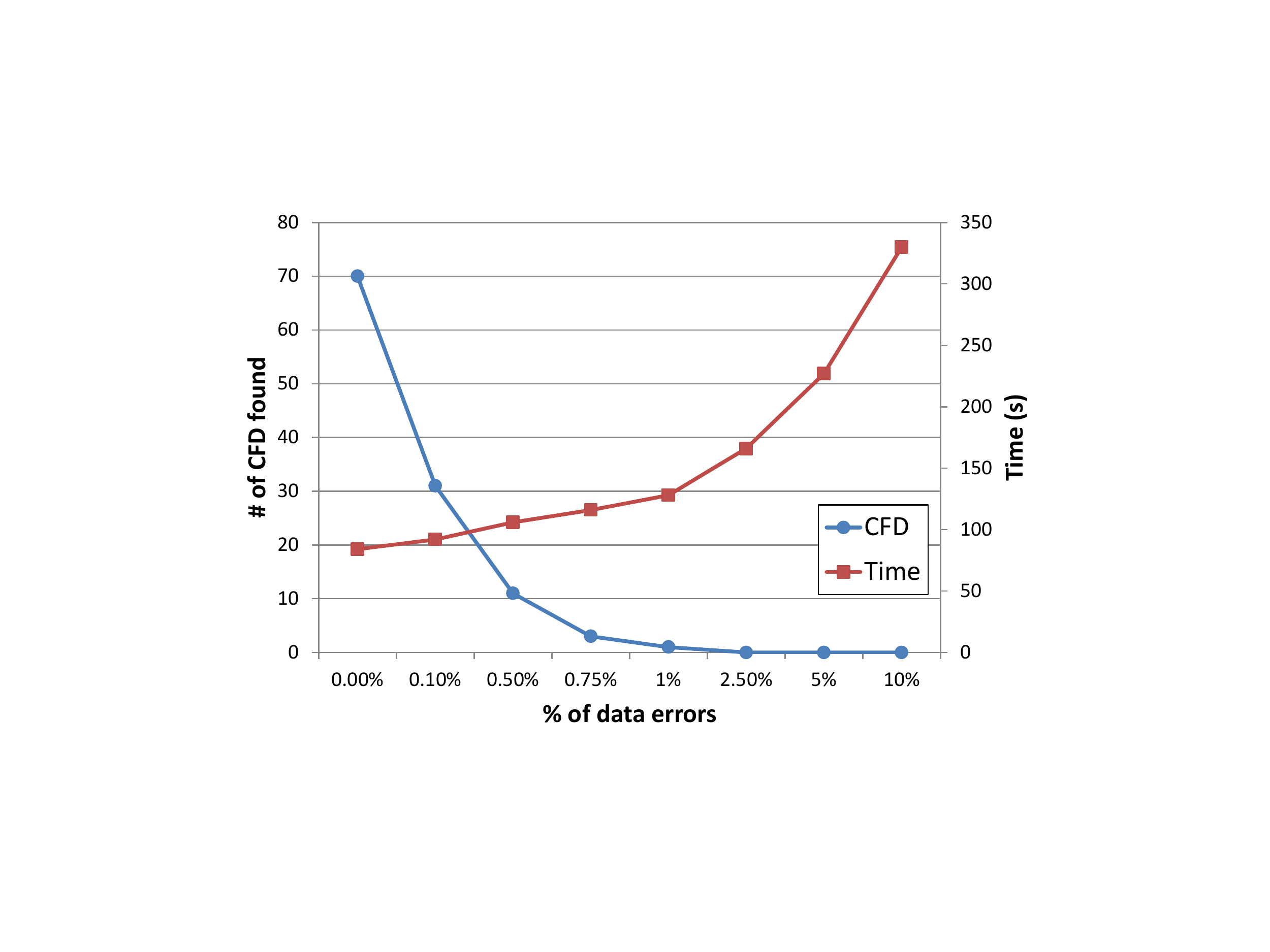}\vspace{-3mm}
\caption{\footnotesize{Learning CFD from dirty web data. It is clear that \#CFD decreases but time increases w.r.t the growth of data errors}}\label{fig:cfd-performance}
\end{figure}

In this section, we present a understanding of how CFD based approaches work on real web data and show their inabilities to clean the web data. As an example, Figure~\ref{fig:cfd-performance} shows the performance of a conditional functional dependency (CFD) learner on the real auto sales data with respect to different levels of noise (e.g., spelling errors, deletion errors, or replacement errors), which are generated randomly. The schema for this dataset is \textsf{car({model, make, car-type, year, condition, drive-train, doors, engine})} and the total number of tuples was over 30,000. For the CFD learner, we directly used the one provided by the authors of \cite{chiang2008discovering}.

Based on the graph, we make one key observation of the deficiency of
CFD: with the growth of percentage of errors in the data, CFD
dramatically finds fewer data editing rules. In specific, with only
1\% errors in the data, the system is unable to learn any rules
As a result, the error data is basically not cleanable or unrepairable. We believe this is mainly due to the fact that: (1) the presence of corrupted values violates possible patterns in the data, making them fractional and inconsistent; (2) On the other hand, finding CFD is deterministic \cite{chiang2008discovering}, i.e., CFD cannot tolerate any errors in the data patterns without any approximations.

We find that the CFD-based methods only work well if the data is
perfectly clean or largely clean (e.g., CFD learner found 70 or 31
rules when data is  100\% or 99.9\% clean as shown in Figure
\ref{fig:cfd-performance}). However, such assumption is rather
unrealistic when we try to clean the real web data since the web is
open and its noise rate would be possibly much higher than any
controlled closed domain (e.g., enterprise database) which itself was
reported to have an average 5\% data errors \cite{redman1998impact}.
Besides, CFD-based approaches are mostly used to make the data
consistent (i.e., data patterns after cleaning tend to conform to
these CFD rules). However, it is not guaranteed that these fixed
errors are the certain errors. To obtain the certain fixes, recent
effort \cite{fan2010towards} suggests to first acquire a clean master
data, learn CFD there and apply the learned rules to clean data.
Unfortunately, while these clean corpus might be easy to establish in
a closed domain, it is hard to do so on the web.

\section{Our Approach}
\label{sec:approach}
The observation mentioned above highlights the importance of developing approaches that can really clean certain errors in dirty web data. In this section, we describe our model and the approach we propose to solving this problem.

\subsection{Conceptual Model}
\label{sec:model}


In this work, we view the data cleaning task as a statistical
inference problem. Let $\mathcal{D} = \{T_1,...,T_n\}$ be the input
dataset. $T_i$ is a tuple with $m$ attributes $\{A_1,...,A_m\}$, which
can be either clean or dirty,  i.e., one or more attributes values are
corrupted. Let $\mathcal{T}^* = \{T_1^*,...,T_n^*\}$ be a correction
candidate set for a tuple $T$. Then, in order to clean $T$, the model
is to find the most likely $T^*$ in $\mathcal{T}^*$  (note that $T^*$ can be as same as $T$ if $T$ is a clean tuple):
\begin{equation}
T^* = \arg\max_{T^* \in \mathcal{T}^*} \textbf{Pr}[T^*|T]
\label{eq:basic}
\end{equation}

In practice, instead of directly optimizing  Equation \ref{eq:basic}, we can solve an equivalent problem by applying Bayes' rule and dropping the constant denominator. So that we have:
\begin{equation}
T^* = \arg\max_{T^* \in \mathcal{T}^*} \textbf{Pr}[T|T^*]\textbf{Pr}[T^*]
\label{eq:basic2}
\end{equation}

\begin{figure}[!h]
\centering\vspace{-2mm}
\includegraphics[trim = 18mm 85mm 18mm 50mm, clip, width=85mm, keepaspectratio]{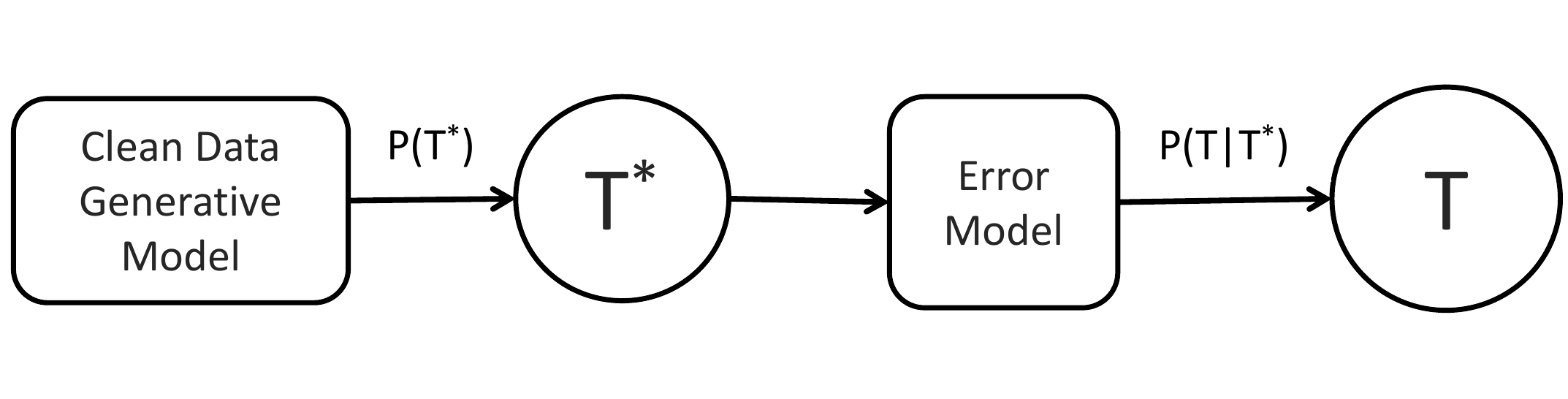}\vspace{-3mm}
\caption{\footnotesize{Conceptual model of our approach}}\label{fig:concept}
\end{figure}

So what is $\textbf{Pr}[T|T^*]$ and $\textbf{Pr}[T^*]$? To answer
this, let us first review how a tuple gets corrupted. We can view
tuples $T$ as being generated by a two-stage process. First, in the generation stage, a (noise free) tuple $T^*$ is generated according to an underlying ``clean" probabilistic data model. Then, this tuple gets corrupted. Which attribute(s) are corrupted is determined by an underlying probabilistic ``error" model and the ``dirty" values for the corrupted attribute(s) are generated from that error model according to some probabilities. The actual representation stored in $\mathcal{D}$ (and seen) is the tuple $T$. Therefore, $\textbf{Pr}[T^*]$ can be viewed as the ``clean" data generative model and $\textbf{Pr}[T|T^*]$ can be viewed as the probabilistic ``error" model. We summarize this error generation process in Figure \ref{fig:concept}.

A benefit of our approach based on generative and error models is its
flexibility. For example, we can make the  error model
accommodate a wide spectrum of possible errors (a common limitation of
current data cleaning approaches is to focus on single type of
errors). Furthermore, one can create an error model to account for
either dependent or independent corruptions.

Now, given the two models, our task is to estimate them. We start with learning the generative model using Bayes networks and later building error model with a maximum entropy model.



\subsection{Data Generative Model}
\label{sec:generativemodel}
Calculating the data generative model $\textbf{Pr}[T^*]$ needs to consider the dependencies between the attributes of possible clean tuple $T^*$. Bayes network seems to be a good choice to model and quantify these correlations.

\begin{figure}[h]
\centering\vspace{-2mm}
\includegraphics[trim = 10mm 10mm 10mm 7mm, clip, width=75mm, keepaspectratio]{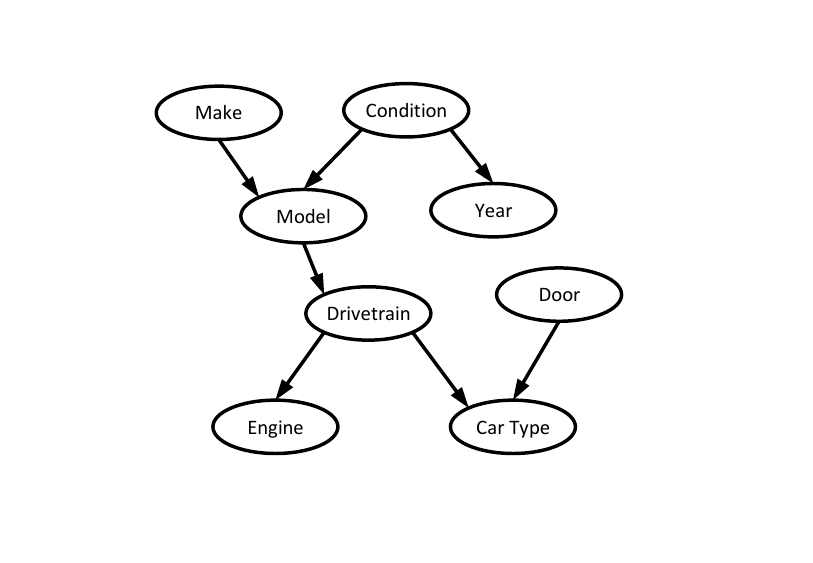}
\caption{\footnotesize{The learned Bayes Network structure of Auto dataset}}\label{fig:bayes-network}
\end{figure}

Learning the Bayes Network usually involves two steps: learning the
topology of the Bayes network and learning its conditional
probability tables (CPTs). For the first step, we use the Bayesian
learning package \emph{Banjo} \cite{hartemink2005banjo} and run it over the
dataset $\mathcal{D}$. Note that although $\mathcal{D}$ may contain
noisy data, but unlike CFD approaches, Bayes network naturally
models the data in a probabilistic way and  thus can tolerate such noise.
Once we have the structure of the Bayes network
$\mathcal{BN}$, we use \emph{Infer.NET} package \cite{InferNET10}  to learn the
parameters (aka conditional probability tables).
The Bayes network thus learned represents $\textbf{Pr}[T^*]$ in a
factored form. In particular, the probability of any specific true
tuple $T^*$ can be read off as a joint probability entry from the
Bayes networks.
%
In Figure~\ref{fig:bayes-network} we show a sample of the Bayes
network structure learned from the auto dataset.

\begin{figure*}[!ht]
\begin{center}
\includegraphics[keepaspectratio, width=400pt, clip, trim=0pt 0pt 0pt 0pt]{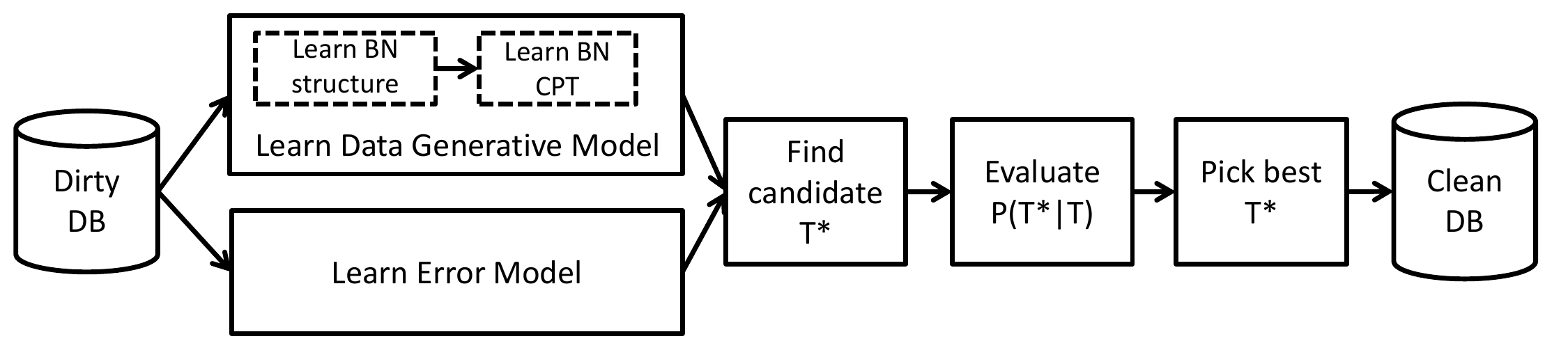}
\caption{The architecture of end-to-end probabilistic web data cleaning system. Our framework requires both data generative model and error model from the raw data. As mentioned in Section \ref{sec:generativemodel}, learning data generative model is based on a two-stage process as depicted in dashed boxes, respectively.}
\vspace{-20pt}
\label{fig-sys-architecture}
\end{center}
\end{figure*}

\subsection{Error Model}
\label{sec:errormodel}

Next, we need to estimate the probabilistic error model
$\textbf{Pr}[T|T^*]$.
To simplify the learning of the error model, we assume that each
attribute is corrupted independently of the other attributes. This
allows us to learn the tuple error model as a product of  the
attribute error models. %
Specifically, we have:
\begin{equation}
\textbf{Pr}[T|T^*] = \prod_{1\leq i \leq m} \mathbf{Pr}[T_{A_i}|T_{A_i}^*]\label{eq:3}
\end{equation}

As mentioned earlier, the error distribution described by our error
model, $\mathbf{Pr}[T_{A_i}|T_{A_i}^*]$, is general enough to
represent any kind of error, as long as the distribution is known. In
this paper, we focus on three types of errors that we observed to be  the
most commonly occurring  in the web data: spelling errors, replacement
errors, and deletion errors. We present different strategies to
characterize them, a summarization is presented in Equation
\ref{eq:4}.

\begin{equation}
 \mathbf{Pr}[T_{A_i}|T_{A_i}^*] = \left\{
  \begin{array}{l l}
    1 & \quad \text{if no error}\\
    f_{ed} (T_{A_i},T_{A_i}^*) & \quad \text{if spelling error}\\
    f_{ds} (T_{A_i},T_{A_i}^*) & \quad \text{otherwise}\\
  \end{array} \right. \label{eq:4}
\end{equation}

The error distribution of a spelling error is based on the
edit-distance feature $f_{ed}$ (see its definition below). Otherwise,
our estimation is based on the distributional similarity feature
$f_{ds}$ (see its definition below). In such a case, the error model $\mathbf{Pr}[T_{A_i}|T_{A_i}^*]$ can be regarded as the probability that one attribute value $T_{A_i}^*$ is replaced by other value $T_{A_i}$. Note also that we can view a deletion error as a special case of the substitution error, i.e., the substituted value is empty (\textsf{NULL} value).

\begin{definition}(Edit-distance feature). This feature $f_{ed}$ is defined based on string edit-distance between two input tuple values. To present it in a probabilistic way, we use the definition in \cite{ristad1998learning}:
\begin{equation}
f_{ed}(T_{A_i},T_{A_i}^*) = \exp \{-ED(T_{A_i},T_{A_i}^*)\}\label{eq:5}
\end{equation}
where $ED(T_{A_i},T_{A_i}^*)$ is the number of edit operations required to transform attribute value $T_{A_i}^*$ into $T_{A_i}$.
\end{definition}

\begin{definition} (Distributional similarity feature). This feature $f_{ds}$ is defined based on the probability of replacing one value with another under a similar context. Formally, we have:
\begin{eqnarray}
f_{ds}(T_{A_i},T_{A_i}^*) = \sum_{c\in C(T_{A_i}, T_{A_i}^*)} \frac{\mathbf{Pr}[c|T_{A_i}^*]\mathbf{Pr}[c|T_{A_i}]\mathbf{Pr}[T_{A_i}]}{\mathbf{Pr}[c]}\label{eq:6}
\end{eqnarray}
where $C(T_{A_i}, T_{A_i}^*)$ is the context of a tuple attribute value, which is a set of attribute values that co-occur with both $T_{A_i}$ and $T_{A_i}^*$. $\mathbf{Pr}[c|T_{A_i}^*] = (\#(c,T_{A_i}^*)+\mu)/\#(T_{A_i}^*)$ is the probability that a context value $c$ appears given the clean attribute $T_{A_i}^*$ in the sample database. Similarly, $P(T_{A_i}) = \#(T_{A_i})/\#tuples$ is the probability that a dirty attribute values appears in the sample database. We calculate $\mathbf{Pr}[c|T_{A_i}]$ and $\mathbf{Pr}[T_{A_i}]$ in the same way. To avoid zero estimates for attribute values that do not appear in the database sample, we use Laplace smoothing factor $\mu$.
\end{definition}


The following example illustrates how distribution similarity between
features is computed.

\begin{example} Consider a tuple $t$: (\textsf{Focus}, \textsf{Honda}, \textsf{JPN}, \textsf{Mid-size}, \textsf{V6}) from group 4 ($g_4$) in Table \ref{tbl:exp2}, where the frequencies are based on the occurrences of certain attribute-values. e.g., 100 tuples (such that they form a group) whose Model=Accord $\wedge$ Make=Honda  $\wedge$ Size=Full-size  $\wedge$ Engine=V6. Based on common knowledge, the value \textsf{Focus} might be \emph{dirty}.\footnote{Focus is well-known Ford car.} There are two possible candidates for the correct value: \textsf{Accord} from $g_1$ or $g_2$, and \textsf{Civic} from $g_3$. To determine which is the right one, we calculate distributional similarity features $f_{ds}$(\textsf{Accord}, \textsf{Focus}) and $f_{ds}$(\textsf{Civic}, \textsf{Focus}).

First, we need to get the context $C(\textsf{Accord}, \textsf{Focus})$. Note that, since there are two groups of \textsf{Accord} car with different engines, the result of their distributional similarity to \textsf{Focus} in $t$ is also different. Nevertheless, let $S_1$ be the set of all the attribute values in the tuples that contain \textsf{Accord} from $g_1$. We have $S_1 = \{\textsf{Honda}, \textsf{JPN}, \textsf{Full-size},\textsf{V6}\}$; Similarly, we have $S_2 = \{\textsf{Honda}, \textsf{JPN}, \textsf{Full-size}, \textsf{V6} \}$, where $S_2$ is the set of co-occurring attribute values of tuples that contain \textsf{Focus} in $g_4$ (since $t$ is from $g_4$). Let the context $C(\textsf{Accord}, \textsf{Focus}) = S_1 \cap S_2 = \{\textsf{Honda},\textsf{JPN},\textsf{Full-size}, \textsf{V6}\}$. Applying Equation \ref{eq:6}, we can get $f_{ds}(\textsf{Accord}, \textsf{Focus}) = 0.179$ conditioned on $g_1$ and $g_4$. Analogously, we can also get $C(\textsf{Civic}, \textsf{Focus}) =\{\textsf{Honda},\textsf{JPN}\}$ and $f_{ds}(\textsf{Civic}, \textsf{Focus}) = 0.082$. As a result, \textsf{Accord} is the right candidate for dirty value \textsf{Focus}. \end{example}\vspace{-8mm}

\begin{table}[h]
\caption{Sample database}
\begin{center}
\begin{small}
\begin{tabular}{ c || c | c | c | c | c || c }
  \hline
  GID & Model & Make & Orig & CarType & Engine &  Freq.\\
  \hline
  $g_1$ & \textsf{Accord}   & \textsf{Honda}    & \textsf{JPN} & \textsf{Full-size} & \textsf{V6} & 100\\
  $g_2$ & \textsf{Accord}    & \textsf{Honda}	& \textsf{JPN} & \textsf{Full-size} & \textsf{V4} & 150\\
  $g_3$ & \textsf{Civic}	& \textsf{Honda}	& \textsf{JPN} & \textsf{Mid-size} & \textsf{V4} & 100\\
  $g_4$ & \textsf{Focus}	& \textsf{Honda}	& \textsf{JPN} & \textsf{Full-size}  & \textsf{V6} & 15\\
  $g_5$ & \textsf{Focus}	& \textsf{Ford }	& \textsf{USA} & \textsf{Compact}  & \textsf{V4} & 105\\
  \hline
\end{tabular}\label{tbl:exp2}
\end{small}
\end{center}\vspace{-2mm}
\end{table}

In practice, we do not know beforehand which kind of error has occurred for a particular attribute. In other words, it is impossible to predict that definitely without knowing the clean version, $T_{A_i}^*$. Furthermore, it is also rather unrealistic to have a single definite error strategy for a given attribute of a tuple. In fact, we want a unified error model which can accommodate all three types of errors (and be flexible enough to accommodate more errors when necessary). For this purpose, we use the well-known maximum entropy framework \cite{berger1996maximum} to leverage all available features, including string edit distance-based feature $f_{ed}$ and distributional-based feature $f_{ds}$. So for each attribute $A_i$, we have our unified error model defined on this attribute as follows:
\begin{eqnarray} \label{eq:7}
\mathbf{Pr}[T_{A_i}|T_{A_i}^*] = \frac{1}{Z}\exp\left\{ \alpha f_{ed}(T_{A_i}^*, T_{A_i}) +  \beta f_{ds}(T_{A_i}^*, T_{A_i}) \right\}
\end{eqnarray}

where $\alpha$ and $\beta$ are the weight of each feature. $Z = \sum_{T^*} \exp \left\{ \sum_{i}  \lambda_i f_{i}(T^*, T)\right\}$ a normalization factor. To compute the entire error model for tuple $T$ and $T^*$, we just plug Equation \ref{eq:7} in Equation \ref{eq:3}. 

\subsection{Putting the Pieces Together}
\label{sec:archicture}
We now describe the working of the system
depicted in Figure~\ref{fig-sys-architecture}. Our system runs as a standalone
application on an offline database which may contain possible corruptions. We first tokenize the entire data and applying the Banjo package to learn the structure of the Bayes network for it. We then provide the learned structure together with the entire database to an inference engine (Infer.NET in our paper) for learning the CPTs. By completing this stage, we have a generative model of the data. In parallel, we define and learn an error model which incorporates three types of errors (call Section \ref{sec:errormodel}). Now we can begin cleaning the database tuple by tuple. For each tuple $T$ in the database, we first find a set of its clean candidate $\mathcal{T}^*=\{T_1^*,...,T_i^*\}$ by looking at all the tuples in the database that are within a certain edit distance of $T$. Then, for each $<T, T^*>$ pair in the database, we now compute the $\mathbf{Pr}[T^*|T]$ value using Equation \ref{eq:basic2} which itself is based two learned models as mentioned above.

Last, we pick the one which maximizes $\mathbf{Pr}[T^*|T]$ and deem it the best $T^*$ and store it as the clean copy of the tuple.

\section{Experiments}
\label{sec:experiments}
In this section, we quantitatively study the performance of our
proposed approach on a large real datasets: Used car sales data. We
present two sets of experiments on evaluating the approach in terms of
(1) the effectiveness and (2) the efficiency.

\subsection{Experimental Setup}
\label{sec:expt-data}
To perform the experiments, we obtained the real data from the web.
The first dataset is \emph{Used car sales} dataset $D_{car}$ which
contains around 10,000 tuples crawled from Google Base. The schema of
this dataset that we used in our experiments was \textsf{car({model,
    make, car-type, year, condition, drive-train, doors, engine})}. We
manually inspected the data to make sure it was clean and deemed the
dataset ``clean". We then introduced three types of noise to
attributes in $D_{car}$. To add noise to an attribute, we randomly
changed it either to a new value which is close in terms of string
edit distance (distance between 1 and 4, simulating spelling errors)
or to a new value which was from the same attribute (simulating
replacement errors) or just delete it (simulating deletion errors).
Such ``dirty" dataset is referred to as ``$D'_{car}$". We used a
parameter $\tau$ ranging from 0.1\% to 5\% for the noise
rate. 

\subsection{Effectiveness}
\label{expt-correctness}
\begin{figure}[t]
\begin{center}
\includegraphics[trim = 55mm 52mm 18mm 50mm, clip, width=100mm, keepaspectratio]{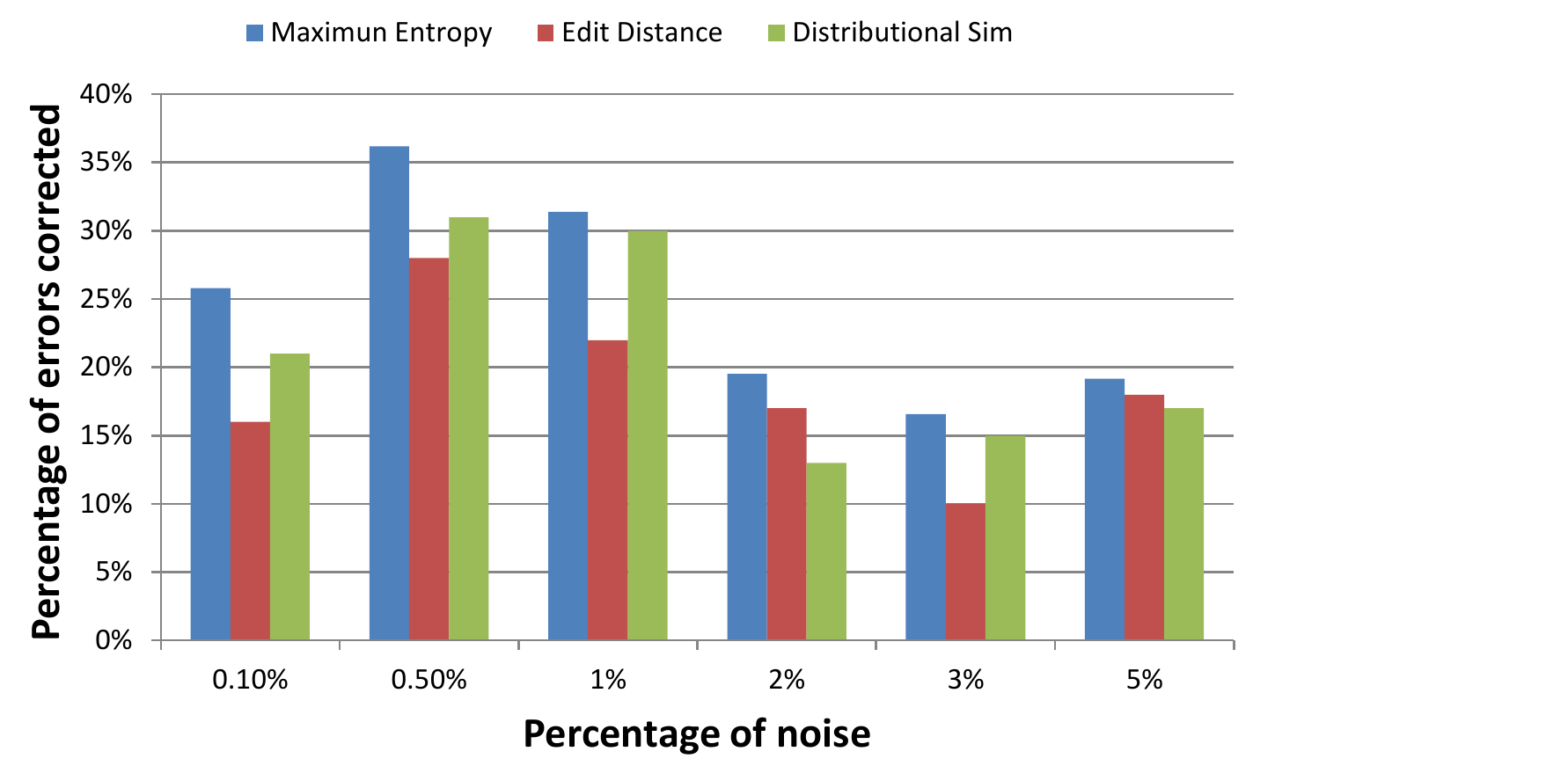}
\caption{The percentage of corrupted values cleaned by the algorithms (using both features, only edit-distance feature, and only distributional similarity feature) as a function of the noise in the database.}
\vspace{-20pt}
\label{fig-corrections-vs-noise}
\end{center}
\end{figure}

\begin{figure}[t]
\begin{center}
\includegraphics[keepaspectratio, width=230pt, clip, trim=0pt 10pt 0pt 0pt]{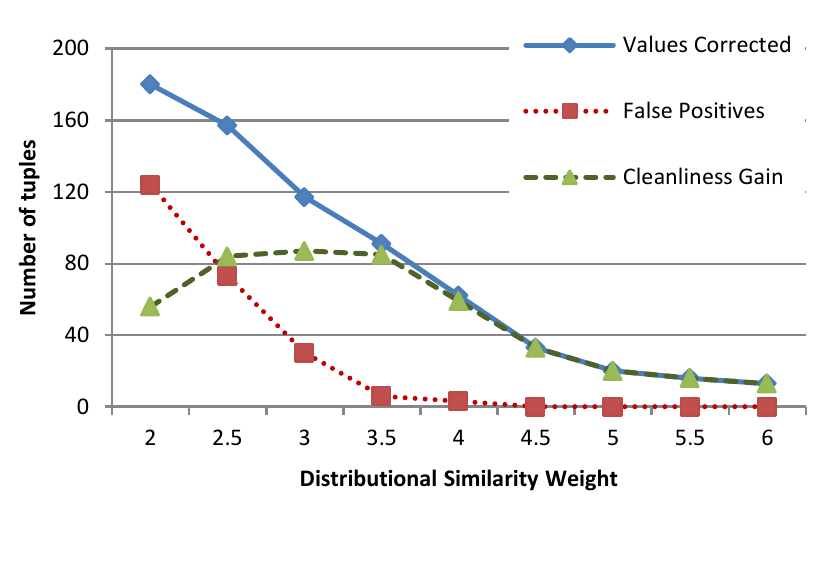}
\caption{The number of values corrected by the algorithm, the number of erroneous values introduced by the algorithm, and the overall increase in the number clean values generated. The x-axis shows the value of the parameter.}
\vspace{-20pt}
\label{fig-gain-vs-parameter}
\end{center}
\end{figure}

We now show the effectiveness of our algorithm in cleaning the noise
data in $D'_{car}$, and demonstrate how the parameters may be varied
to obtain the desired results. In
Figure~\ref{fig-corrections-vs-noise} we show the resilience of the
algorithm to noise in the input database $D'_{car}$. The weight ($\alpha$) for the string edit distance feature $f_{ed}$ was fixed at 2.3 while the weight ($\beta$) for the distributional similarity feature $f_{ds}$ was fixed at 3.5.
These values were chosen based on the results from
Figure~\ref{fig-gain-vs-parameter}, which we explain in the next
paragraph. In addition, to evaluate the effectiveness of the maximum entropy model that we adopted, we compare its cleaning performance with the ones obtained by the cleaning algorithms that use one single type of features at a time (in other words, we set $\alpha=0$ and  $\beta=0$ in Equation \ref{eq:7} respectively and get accordingly results). As we can see from this figure, all algorithms achieve substantial reduction in the noise of the data. Specifically, at 1\%
noise in the data, our algorithm which leverage all features corrects more than 31\% errors in the
data, whereas CFD based methods failed to find even a single CFD (call
Figure \ref{fig:cfd-performance}) and are thus not able to fix any data corruptions. The number
of false positives for each of these cases was less than or equal to
11 tuples (which is a very small percentage of the corpus size). Besides, it is clear that using the maximum entropy model from to combine all features achieves better results than using them alone.\newline

\noindent
{\bf Setting $\beta$:} Recall that in our approach we have two weights that can be adjusted: the weight given to the distributional similarity ($\beta$), and the weight given to the edit distance  ($\alpha$). The ratio of these two weights depends on which kind of error is more likely to occur. We found that setting the edit distance weight to $0.667  \times \beta$ yields the best results. Keeping this ratio fixed, in Figure~\ref{fig-gain-vs-parameter}, we show how the algorithm performs as $\beta$ is changed. The ``values corrected'' data points in the graph correspond to the number of attribute values that were erroneous in the input data that the algorithm successfully corrected (when checked against the ground truth).

The ``false positives'' are the number of legitimate values that the algorithm changed to an erroneous value. When cleaning the data, our algorithm chooses a candidate tuple based on both the prior of the candidate as well as the likelihood of the correction given the evidence. Low values of the parameter $\beta$ give a higher weight to the prior than the likelihood. In other words, a lower value of the parameter indicates a higher likelihood of changing the tuple. As a result, some legitimate tuples are ``corrected" to a tuple that has a much larger prior. As $\beta$ is increased, the number of such false positives reduces. However, this also reduces the number of values corrected, because some kinds of unlikely errors no longer justify the higher cost of correction.

The ``overall gain'' in the number of clean values is calculated as the difference of clean values between the output and input of the algorithm. In this particular experiment, there were 357 errors in the input data, of which the best correction was obtained at a parameter value of 3.0, where the overall gain was 87 clean values.

\subsection{Efficiency}
\begin{figure}[!ht] \vspace{-3mm}
\centering\hspace{-12mm}
\subfloat[Time vs. \#Tuples]{
\includegraphics[keepaspectratio, width=50mm, clip, trim=0pt 10pt 0pt 15pt]{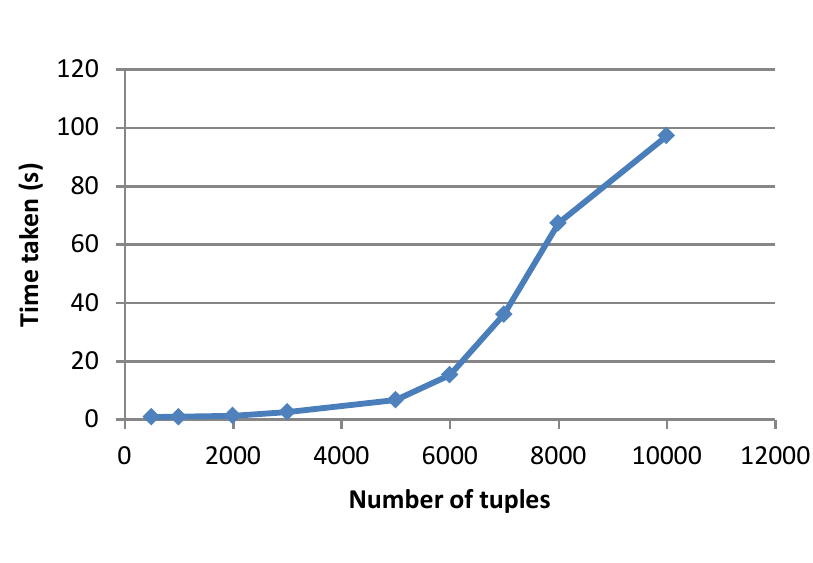}
\label{fig-time-vs-tuples}}
\,\hspace{-4mm}
\subfloat[Time vs. \%Noise]{
\includegraphics[keepaspectratio, width=50mm, clip, trim=0pt 10pt 0pt 15pt]{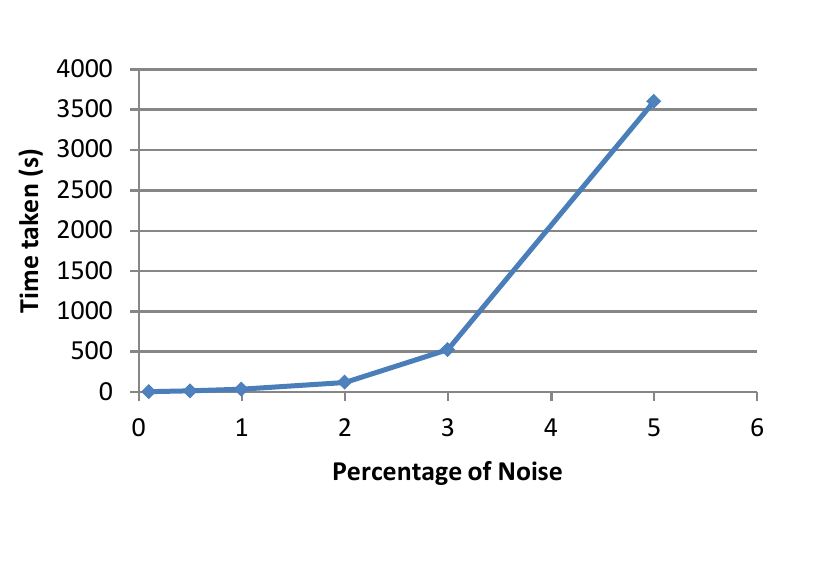}
\label{fig-time-vs-noise}\hspace{-12mm}
}\vspace{-1mm}
\caption{Time taken by the algorithm to clean the database. In (a) we fixed with 0.5\% data noise; In (b) we fixed \#Tuples=5k.}\vspace{-1.5mm}\label{time=noise}
\end{figure}

In Figure~\ref{fig-time-vs-tuples} and Figure \ref{fig-time-vs-noise} we show the time taken by the algorithm (using Maximum entropy. This graph includes both the time taken to learn the generative model as well as the time taken to clean every tuple of the database. As can be seen, the algorithm completes in a reasonable amount of time, even with 10,000 tuples in the database.\vspace{3mm}

We show the effect of noise on the time taken by the algorithm in Figure~\ref{fig-time-vs-noise}. For this curve, the number of tuples was kept constant at 5,000 tuples. As can be seen from the figure, the time taken by the algorithm increases as the percentage of noise in the data increases. This is because for every tuple that we have to clean, we have a much larger set of candidate $T^*$s to consider. Adding noise to the dataset effectively increases the number of different tuples within the edit distance threshold of the data, thus a much larger number of error model comparisons need to be made.


\section{Conclusion}
\label{sec:conclusion}

In this paper, we focused on approaches for cleaning inconsistent web
data. We showed that the current state of the
  art approaches, which  learn and use
  conditional functional dependencies (CFDs) to rectify data, do not
  work well with web data as they demand clean master data for
  training. We proposed  a fully
  probabilistic framework for cleaning data that involved
  learning both the generative and error (corruption) models of the
  data and using them to clean the data. For generative models, we
  learn Bayes networks from the data. For error models, we consider a
  maximum entropy framework for combing multiple error processes.
 The generative and error
  models are learned directly from the noisy data. Preliminary
  experimental results on web data showed that our probabilistic
  approach is able to reduce errors in the data long after CFD-based
  methods fail to be effective. 

\bibliographystyle{abbrv}
\bibliography{kydd}

\end{document}